\documentclass[
reprint,
amsmath,amssymb,
aps,
prl,
]{revtex4-2}

\usepackage{float}
\usepackage{mathtools}
\usepackage{graphicx}% Include figure files
\usepackage{dcolumn}% Align table columns on decimal point
\usepackage{bm}% bold math
\usepackage{upgreek}
\usepackage{lipsum, babel, xcolor}
\usepackage{amsmath, amssymb}
\usepackage{comment}
\usepackage{natbib}
\usepackage{physics}
\usepackage{hyperref}
\usepackage{multirow}
\hypersetup{colorlinks=true,citecolor=blue}
\usepackage{booktabs}
\usepackage{siunitx}
\usepackage{tabularx}
\usepackage{mathptmx}

\begin{document}

% \preprint{APS/123-QED}

% \title{Optical and DC stability of microring and coupling resonators on thin-film lithium tantalate at high powers}
\title{High-power handling and bias stability of thin-film Lithium Tantalate microring resonators and coupling modulators}

\author{Ayed Al Sayem, Shiekh Zia Uddin, Ting-Chen Hu, Alaric Tate, Mark Cappuzzo, Rose Kopf, Mark Earnshaw}

% \author{Ayed Al Sayem$\mathrm{^{1}}$, Shiekh Uddin$\mathrm{^{1}}$, Ting-Chen Hu$\mathrm{^{1}}$, Alaric Tate$\mathrm{^{1}}$, Mark Cappuzzo$\mathrm{^{1}}$, Rose Kopf$\mathrm{^{1}}$, Mark Earnshaw$\mathrm{^{1}}$}
% \author{Mark Earnshaw$\mathrm{^{1}}$}
%author{Second Author}%
%\email{ayed.sayem@nokia-bell-labs.com}
% \author[1,*]{Ayed Al Sayem}
% \author[1]{Heqing Huang}
% \author[1]{Ting-Chen Hu}
% \author[1]{Alaric Tate}
% \author[1]{Mark Cappuzzo}
% \author[1]{Rose Kopf}
% \author[1]{Mark Earnshaw}
\affiliation{%
  Nokia Bell Labs, Murray Hill, NJ, USA 
}%
\date{\today}% It is always \today, today,

\begin{abstract}
In this paper, we demonstrate ultra-high power-handling capability and DC-bias stability of optical microring and electro-optic (EO) coupling resonators on the thin-film lithium tantalate (TFLT) platform. We show that, with annealing, oxide cladded TFLT microring resonators can handle several Watts ($\sim4$\,W) of circulating power with minimal frequency shift ($<1$\,GHz) and no observable photo-refractive effect. Furthermore, we demonstrate a compact $2$\,mm coupling modulator achieving a low $V_{\pi}$ of $3$\,V with stable bias and phase control in the telecom C-band.
\end{abstract}

\maketitle

\section{Introduction}

%% papers to cite
% TFLN Review: 
% Di Zhu \cite{zhu2021integrated}

% Photo-refractive effect TFLN/TFLT:
% \cite{yan2020tantalate_microdisk}
% \cite{holtmann2004photorefractive}
% \cite{xu2021mitigating}

% Photo-refractive effect TFLN: Device instability:
% \cite{lu2021ultralow} %% my paper
% \cite{Ahmed2025Universal}

% Non-linear optics:
% Frequency comb:
% \cite{zhang2019broadband}
% \cite{wang2019monolithic}
% \cite{gong2020near}

% parametric generation:
% \cite{lu2021ultralow}
% \cite{sayem2021efficient} %% my paper
% \cite{guo2023ultrafast}
% \cite{nehra2022few}

% Quantum photonics:
% Source:
% \cite{kundu2024periodically} %% my paper
% \cite{xin2022spectrally}

% superconducting detector:
% \cite{sayem2020lithium} %% my paper
% \cite{lomonte2021single}
% \cite{colangelo2024molybdenum}

% cryogenic:
% \cite{sayem2020lithium} %% my paper
% %% m2o
% \cite{xu2021bidirectional} %% my paper
% \cite{McKenna2020Cryogenic}
% \cite{Holzgrafe2020Cavity}

% TFLT papers:

% Kippenberg:
% \cite{wang2024lithium} % general nature
% \cite{zhang2025ultrabroadband} % modulation 
% \cite{wang2024ultrabroadband}

% Loncar:
% \cite{}

% Coupling modulator papers
% \cite{xue2022breaking}
% \cite{al2025multi} %% my paper

% Key TFLN modulator papers:
% \cite{xu2020high}
% \cite{kharel2021breaking}
% \cite{xu2022dual}

% PPLN source:
% \cite{ma2020ultrabright}
% \cite{kundu2024periodically} %% my paper

Thin-film lithium niobate (TFLN) has been developed over the past decade and is currently one of the primary candidate platforms for next-generation optical communication networks, both short-haul and long-haul \cite{zhu2021integrated}. While TFLN fabrication techniques have significantly matured, the platform is strongly affected by the photorefractive (PR) effect, which severely limits power handling capability and long-term device stability, thereby limiting practical applications. This inherent instability remains a critical barrier to the wide adoption of TFLN, as it necessitates strictly limited optical power levels and complicates the achievement of reliable, steady-state operation in practical environments \cite{xu2021mitigating,xu2021bidirectional,lu2021ultralow,sun2017nonlinear}. Although ultra-efficient photonic devices on the TFLN platform have been demonstrated, such as ultra-low threshold parametric oscillator \cite{lu2021ultralow}, photon-pair generation \cite{ma2020ultrabright}, microwave to optical quantum transducer \cite{xu2021bidirectional,McKenna2020Cryogenic,Holzgrafe2020Cavity}, low-drive ultra high speed modulator \cite{xue2022breaking}, instability, and drift due to the PR effect fundamentally limits device performance \cite{xu2021mitigating, Arge2025SqueezedTFLN}. Normalized efficiency is exceptionally high for many of the demonstrations on the TFLN platform, but the inability to increase on-chip optical power directly hinders practical applications. The PR effect also causes DC bias instability \cite{xu2020high,Warner2025DCStable}, which makes stable low-power bias circuits not possible on TFLN \cite{xu2020high}. The thermo-optic (TO) effect can be used for bias control for traveling wave-modulator \cite{xu2020high}, or resonators \cite{al2025multi}.  However, as the TO effect is weak for the LN platform, it causes very high static power consumption \cite{xu2020high}. This then prevents scaling to practical, large-scale, photonic circuits on TFLN, where many such bias circuits are necessary. Especially for quantum applications where cryogenic stable bias circuits are required \cite{psiquantum2025manufacturable}, TFLN loses its fundamental advantage of the EO effect due to this instability \cite{xu2021bidirectional,Holzgrafe2020Cavity,McKenna2020Cryogenic}.  

Lithium tantalate (LT) has very similar electro-optic (EO) properties but offers much higher optical damage threshold than LN \cite{yan2020tantalate_microdisk,holtmann2004photorefractive,althoff1990linbo3,kong2020optical_damage}. Some early demonstrations have shown very promising results on the TFLT platform such as low propagation loss \cite{wang2024lithium}, high-rate data modulation \cite{wang2024ultrabroadband,niels2025high}, frequency comb generation \cite{zhang2025ultrabroadband} etc. The open question is whether TFLT can perform with similar efficiency as TFLN but with proper stable operation, which can pave the way from proof-of-principle demonstrations to real-world applications.

In this article, we investigate the high-power dynamics of optical microring resonators on the TFLT platform at circulating powers up to $4$\,W. We show that even at these extreme intensities, TO effect remains dominant over the PR effect. Crucially, we find that annealing suppresses the PR effect to the point that it becomes unobservable, even with an already weak TO response.  We show that TFLT has orders of magnitude better optical power handling than TFLN, AlGaAs, and $\mathrm{Ta_{2}O_{5}}$ but slightly worse than SiN.  By exploiting the unique power-handling capabilities of TFLT, we report the platform's first resonant coupling modulator. The device exhibits an exceptionally enhanced modulation efficiency ($V_\pi L = 0.6\text{ V}\cdot\text{cm}$), requiring just $3$\,V $V_\pi$ for a $2$\,mm interaction length.  we also study the stability of a coupling modulator on the TFLT platfrom which is a direct application for high-speed optical communications \cite{xue2022breaking, ChenMistry2024JLT}. We then show stable bias and phase control of the coupling modulator, thereby enabling practical utilization for ultra-high speed and low energy optical communication with stable resonant EO modulators.

% For a complete picture, we also study the stability of a coupling modulator on the TFLT platfrom which is a direct application for high-speed optical communications \cite{xue2022breaking, ChenMistry2024JLT}. This is the first demonstration of a coupling modulator on the TFLT platform and we show exceptional enhanced modulation efficiency of just $3$\,V $V_\pi$ with only $2$\,mm long modulation section ($0.6$\,Vcm $V_\pi L$). We show stable bias and phase control of the coupling modulator, thereby enabling practical utilization for ultra-high speed and low energy optical communication with stable resonant EO modulators.

\section{PR effect on TFLT microring resonator}

\begin{figure*}[ht]
    \centering
    \includegraphics[width = 0.95\textwidth]{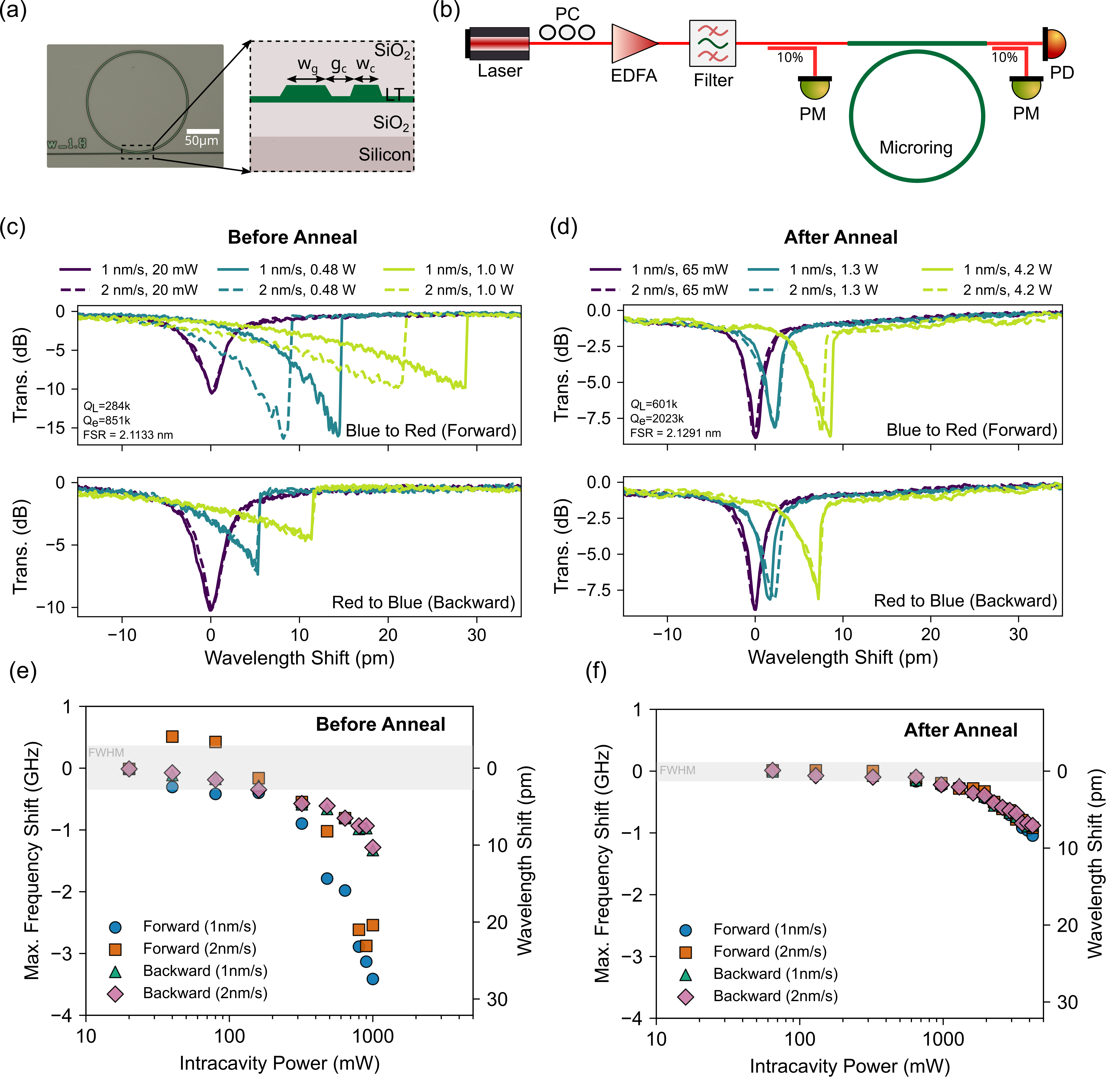}
    \caption{\textbf{Device architecture and high-power characterization of the TFLT micro-ring resonator:} (a) Optical microscope image of the fabricated thin-film lithium tantalate (TFLT) micro-ring resonator ($50\text{ }\mu\text{m}$ scale bar). The inset shows a cross-sectional schematic of the coupling section, highlighting the etched rib waveguide structure on a silicon substrate. Here, $\mathrm{w_{g}=1.8\,\mu m, g_{c}=0.75\,\mu m, w_{c}=0.8\,\mu m}$
(b) Schematic of the high-power measurement setup, incorporating an Erbium-Doped Fiber Amplifier (EDFA) and a tunable filter to characterize the resonator's nonlinear response.
(c–d) Power-dependent transmission spectra before (c) and after (d) thermal annealing. The spectra compare forward (blue-to-red) and backward (red-to-blue) wavelength scans at various speeds ($1\text{ nm/s}$ and $2\text{ nm/s}$). Post-annealing results show significantly reduced resonance distortion and thermal bistability.
(e–f) Extracted maximum resonance frequency shift as a function of intracavity power before (e) and after (f) annealing. The annealed device demonstrates superior power handling, with resonance shifts remaining within the full-width at half-maximum (FWHM) for intracavity powers up to approximately $1\text{ W}$.}
    \label{Fig1}
\end{figure*}

Fig.\ref{Fig1}(a) shows the optical image of the micro-ring resonator used in this paper. The cross-section of the micro-ring resonator in the coupling section is shown in the inset of Fig.\ref{Fig1}(a). Device fabrication begins with a 4-inch TFLT wafer on a silicon substrate with a $\mathrm{4.7\mu m}$ Box oxide, commercially available from NanoLN. Photonic devices are photo-lithographically defined and subsequently etched using $\mathrm{Ar^{+}}$ plasma. After etching the optical device layer, $\mathrm{1.5\,\mu m}$ thick oxide is deposited using plasma-enhanced chemical vapor deposition (PECVD). Co-planar waveguide electrode layers are defined by photolithography and subsequent etching or lift-off process for the coupling resonator. The wafer was diced, and the individual chips were polished for efficient fiber-to-chip coupling.
% \begin{figure*}[ht!]
%     \centering
%     \includegraphics[width = 0.85\textwidth]{Figures/Fig3.png}
%     \caption{(a) Schematic of the measurement setup for the coupling modulator. PC: polarization controller, PM: power meter, AWG: arbitrary wave generator, PD: photo-detector. G: ground, S: signal. (b) Transmission of the coupling modulator as a function of wavelength under different bias voltages. (c) Transmission spectra as a function of time at different bias voltages.}
%     \label{Fig2}
% \end{figure*}
\begin{figure*}[ht!]
    \centering
    \includegraphics[width = 0.95\textwidth]{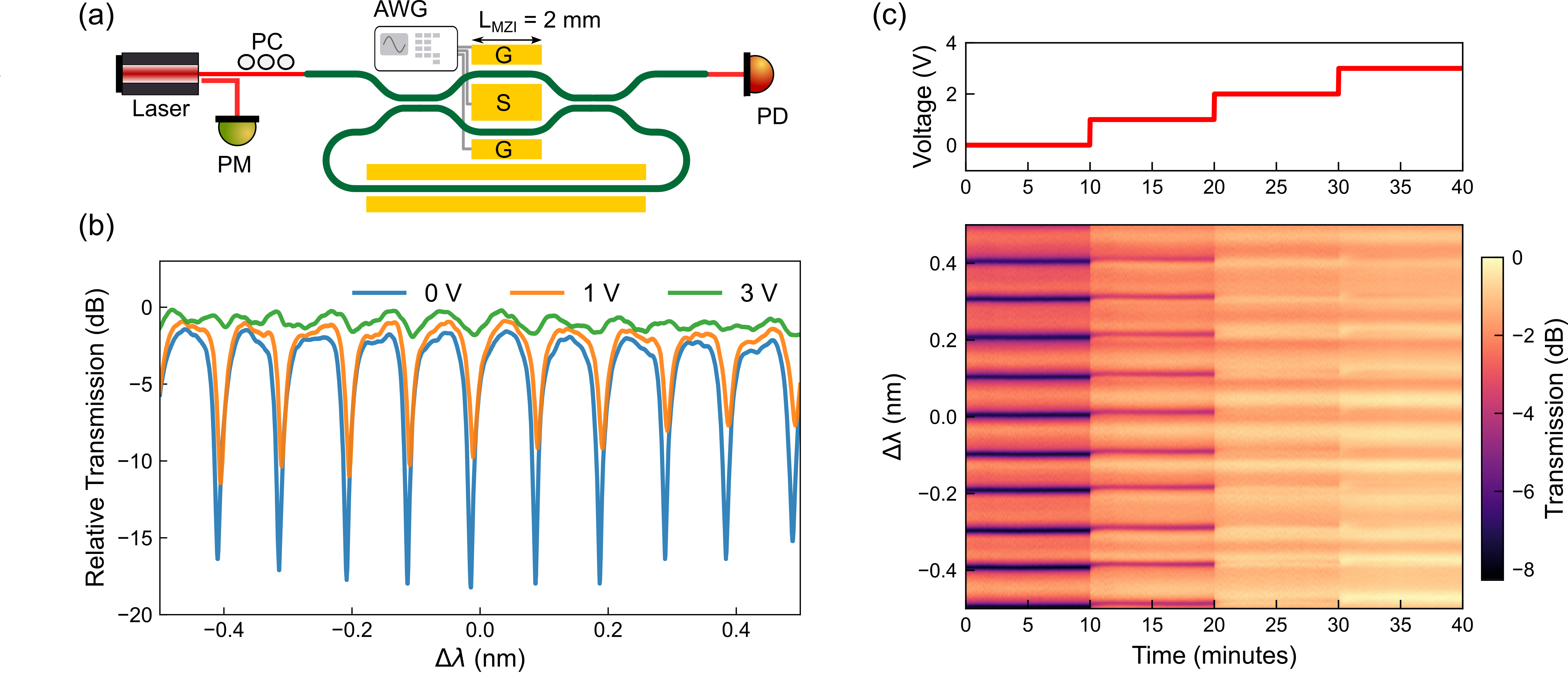}
    \caption{ \textbf{Characterization of the TFLT coupling modulator:} (a) Schematic of the experimental setup for characterizing the device, including a laser source, polarization controller (PC), and power meter (PM). The arbitrary waveform generator (AWG) provides voltage to the Signal (S) and Ground (G) electrodes of the coupling modulator, which has an active length of $2\text{ mm}$ with a $\mathrm{5.5\,\mu m}$ gap between the ground and signal electrode.
(b) Relative transmission spectra at bias voltages of $0\text{ V}$, $1\text{ V}$, and $3\text{ V}$. 
(c) Top: Applied DC bias voltage stepped from $0\text{ V}$ to $3\text{ V}$ in $1\text{ V}$ increments every 10 minutes. Bottom: Time-resolved transmission heatmap showing the spectral response over a 40-minute duration.} 
% The consistent intensity levels within each voltage step highlight the device’s operational stability and the predictable modulation of transmission levels
    \label{Fig2}
\end{figure*}

Fig.\ref{Fig1}(b) shows the measurement setup. Light from a tunable laser (Santec-570) is first amplified by an erbium-doped fiber amplifier (EDFA) and filtered by a bandpass filter with a bandwidth of $10$\,nm. Output light from the filter is then coupled to the DUT using a lensed fiber. A fiber-based splitter is used before the DUT to monitor the launched optical power in a slow optical power meter (PM). A polarization controller (PC) is used to optimize the polarization for the TE mode. After passing through the device, the light is coupled to another lens fiber. A second fiber splitter is used to monitor the output power from the device using another optical PM. The other output of the fiber splitter is sent to a fast photo-detector (PD).

\begin{figure*}[ht]
    \centering
    \includegraphics[width = 0.95\textwidth]{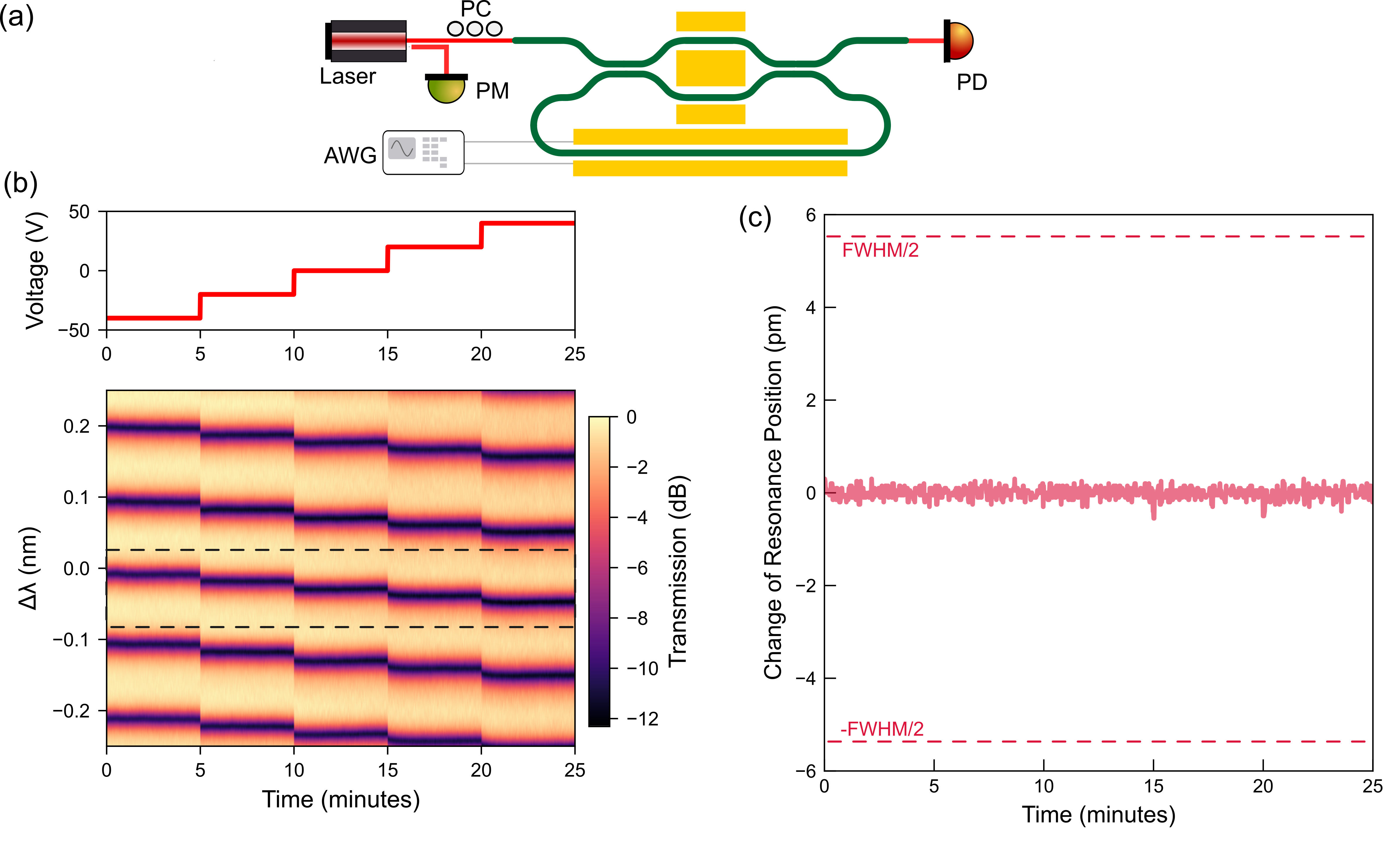}
    \caption{\textbf{Phase stability and repeatability:} (a) Schematic of the experimental setup used for phase stability measurements, featuring a laser source, polarization controller (PC), power meter (PM), and an arbitrary waveform generator (AWG) to drive the electrodes, with the output captured by a photodetector (PD).
(b) Top: Applied phase bias voltage applied in discrete steps. Bottom: Corresponding transmission spectra over time, showing the resonance shift as a function of the stepped voltage.
(c) Stability analysis showing the change in resonance position over 25 minutes. The resonance shift remains below $0.1 \text{ pm}$, well within the half-maximum (FWHM) boundaries.}

% (d) Top: Application of a triangular voltage waveform. Bottom: Time-resolved transmission spectra demonstrating continuous and predictable resonance tuning.
% (e) The standard deviation of resonance positions at specific voltages during the triangular sweep remains below $1 \text{ pm}$, significantly lower than the resonance FWHM.}
    \label{Fig3}
\end{figure*}
We first measure the as-fabricated microring resonators with PECVD oxide cladding and without any annealing. Oxide cladding is necessary for proper velocity matching to achieve high-speed modulation \cite{zhang2021integrated}, hence we study the optical power handling only with oxide cladding. In Fig.\ref{Fig1}(c), we plot the transmission from the device as a function of wavelength. Here, we scan in both forward (blue to red) and backward (red to blue) directions with different scan speeds. Material platforms such as LN or LT, which have a PR effect, can show distinctive behavior with respect to scan direction as the PR effect blue shifts the resonance, whereas the thermo-optic (TO) effect red shifts the resonance \cite{xu2021mitigating}. With forward scanning, the strong PR effect also reduces the linewidth of the resonance, making the fitted quality factor, $\mathrm{Q_{L}}$, artificially high. Such effects have been observed to be quite drastic for TFLN \cite{xu2021mitigating}. As can be observed from Fig.\ref{Fig1}(c), forward and backward scans show different dynamics. With forward scanning, we observe a larger frequency shift than with backward scanning. This TFLT behavior is the opposite to that of TFLN \cite{xu2021mitigating} and confirms the PR effect is weaker than the TO effect in TFLT. In Fig.\ref{Fig1}(e), we show the maximum frequency shift as a function of intra-cavity power before annealing. With 1\,W of intra-cavity power, we observe a maximum frequency shift of $\mathrm{\sim3.5}$\,GHz. We then anneal the device at $\mathrm{500^{o}C}$ for 2\,hours and redo the measurements. After annealing, we observe a negligible difference in the resonance shape and a similar frequency shift for both forward and backward scanning, which can be observed from Fig.\ref{Fig1}(d). The quality factor of the devices also increases by a factor of $\mathrm{\sim2}$ after annealing. In Fig.\ref{Fig1}(f), we show the maximum frequency shift as a function of intra-cavity power after annealing. The frequency shift decreases from $\mathrm{\sim3.5\,GHz}$ to $\mathrm{\sim160\,MHz}$ at $1$\,W of intra-cavity power. With $\mathrm{\sim4\,W}$ of intra-cavity power, we only observe $\mathrm{\sim1\,GHz}$ frequency shift. There is also only a small difference in the frequency shift for forward and backward scanning at any power level, indicating minimal PR effect. We compare the rate of normalized TO-induced frequency shift as a function of energy density of a microring resonator for different material platforms in Table \ref{tab1}. Similar to Ref. \cite{gao2022probing}, we use energy density as it is independent of the device geometry and quality factors which vary for different material platforms. From Table \ref{tab1}, we see that the TFLT platform performs better than TFLN by $\sim9$\, order of magnitude. Only SiN performs better in terms of TO frequency shift than TFLT. The details of the theoretical calculation of the resonance frequency shift incuding photo-refractive and photo-thermal effect is provided in the supplementary information.

\begin{table}[htbp]
\centering
\caption{Photo-thermal Stability Comparison}
\label{tab1}
\footnotesize % Reduces font size slightly to fit column
\begin{tabularx}{\columnwidth}{@{} l S[table-format=1.1e-1] l @{}} 
\toprule
\textbf{Material} & {\textbf{Stability Factor$^\ast$}} & \textbf{Ref.} \\
& {(\unit{\milli\joule^{-1}\centi\meter^3})} & \\
\midrule
Si$_3$N$_4$            & 2.6e-8  & \cite{gao2022probing} \\ 
Si                     & 1.0e-4  & \cite{de2019power} \\ 
Ta$_2$O$_5$            & 8.8e-7  & \cite{gao2022probing} \\ 
Al$_{0.2}$Ga$_{0.8}$As & 2.1e-6  & \cite{gao2022probing} \\ \addlinespace
TFLN (Oxide Coated)        & 4.2e2   & \cite{xu2021mitigating} \\ \addlinespace
\textbf{TFLT (Before Anneal)}  & 9.5e-7  & \textbf{This Work} \\
\textbf{TFLT (After Anneal)}   & 1.8e-7  & \textbf{This Work} \\
\bottomrule
\\
\multicolumn{3}{l}{$^\ast$Defined as $\frac{1}{\omega_0}\frac{\partial\delta\omega_m}{\partial\rho}$, where $\delta\omega_m$ is the resonance frequency shift,} \\
\multicolumn{3}{l}{$\rho$ is the intracavity energy density, and $\omega_0$ is the resonance frequency.} \\
\end{tabularx}
% \label{tab1}
\end{table}

\section{Bias-stabile operation of coupling modulators}
Conventional traveling wave electro-optic modulators on the TFLN and TFLT platforms are primarily limited in drive voltage and RO bandwidth \cite{kharel2021breaking,zhang2021integrated} by RF loss from the electrodes \cite{zhang2021integrated, wang2024ultrabroadband}. Ring modulators where the ring resonant wavelength is tuned have been shown to enhance modulation efficiency but are limited by photon lifetime. Coupling modulators modulate the resonator coupling strength so the photon-lifetime is directly modulated \cite{sacher2008dynamics}. Thus, the EO bandwidth of such a modulator is no longer limited by the photon-lifetime \cite{sacher2013coupling, xue2022breaking, ChenMistry2024JLT}. Coupling modulators can thus break the voltage-bandwidth constraint and have been demonstrated on the TFLN platform \cite{xue2022breaking,al2025multi}. Although coupling modulators have been demonstrated on the TFLN platform \cite{xue2022breaking, al2025multi}, the large PR effect \cite{xu2021mitigating,sun2017nonlinear} prevents use in practical applications where high power (eg: $10-100$ mW used in modern communication) and, most importantly, stable operation is necessary. Here, we show a coupling resonator/modulator on the TFLT platform for the first time and study the bias and phase stability of the EO modulation as a function of time. 
Fig.\ref{Fig2} shows the measurement setup for the coupling modulator. Light from a fast-scanning laser (Santec-570) is coupled to the DUT using a lens fiber. A polarization controller is used to launch TE mode only. An arbitrary waveform generator (AWG) is used to drive the modulator at different bias voltages. The length of the modulator electrodes is $\mathrm{L_{MZI}=2\,mm}$ and the gap between the signal and electrode is $5.5\,\mu m$. The directional couplers of the coupling modulators are designed for a $50:50$ splitting ratio near the C-band. The details of the coupling modulator design and system-level measurement results will be presented in a separate work. In this paper, we focus on the bias stability. Fig.\ref{Fig2}(b) shows the transmission from the device as a function of wavelength when different bias voltages are applied. At $\mathrm{V_{bias}=0\,V}$, the resonator is critically coupled, showing a dip in the transmission. At $\mathrm{V_{bias}=3\,V}$, the coupling condition changes from a critical coupled to extremely uncoupled (no-coupling) with an extinction of more than 15\,dB and with minimal frequency shift. The equivalent V$_\pi$ of a straight waveguide modulator is 14\,V, hence the V$_\pi$ is $\mathrm{\sim 4.7\, times}$ improved. Further enhancement can be achieved with a higher optical quality factor, which can be achieved either through annealing or with a more effective dry and wet etching recipe. In Fig.\ref{Fig2}(c), we plot the transmission of the coupling modulator as a function of time for different bias voltages. We observe no major shift in transmission at different bias voltages, indicating ultra-stable operation. In Fig.\ref{Fig3}(b) and Fig.\ref{Fig3}(c), we show the phase repeatability of the resonator itself as a function of the resonator bias instead of the coupling bias as shown in Fig.\ref{Fig3}(a). We apply a high voltage to the resonator arm, ranging from $-50$\,V to $50$\,V, and measure the transmission from the device as a function of time. We observe negligible transmission drift as a function of time, even with high DC voltages, as can be observed from Fig.\ref{Fig3}(b) and Fig.\ref{Fig3}(c).  

\section{Conclusion}
In conclusion, we demonstrate that with proper annealing, oxide-cladded photonic devices such as microring resonators on the TFLT platform can handle watt-level power with minimal degradation due to the PR effect. We also demonstrate the first coupling modulator on the TFLT platform with stable bias and a $V_\pi$ of just 3\,V with a short 2\,mm modulation electrode. These findings will pave the way for many practical classical and quantum devices on the TFLT platform.

\section{Author contribution}
A.S. designed the photonic devices and developed the fabrication process flow with T.H, A.T, M.C., and R. K.. T.H, A.T, M.C., and R. K. fabricated the devices. A.S and S.U performed the measurements. A.S, S.U wrote the paper with technical feedback from M.E.

\section{Funding}
Nokia Corporation of America.

\bibliography{Reference}

% \include{attached_supp.tex}

% \onecolumngrid
% \input{Supplementary}

\end{document}